%% file: main.tex
\documentclass[journal]{IEEEtran}
\usepackage{amsmath,amssymb,amsthm}
\usepackage{algorithmic}
\usepackage{algorithm}
\usepackage{array}
\usepackage[caption=false]{subfig}
\usepackage{textcomp}
\usepackage{stfloats}
\usepackage{url}
\usepackage{verbatim}
\usepackage{graphicx}
\usepackage[hidelinks,colorlinks=true,linkcolor=blue,citecolor=blue,urlcolor=blue]{hyperref}
\usepackage[nolist]{acronym}
\usepackage{lipsum}  
\usepackage{enumitem}  
\usepackage{tabularx}
\usepackage{booktabs}
\usepackage{setspace}
\usepackage{relsize}

\newcommand\dunderline[2][.4pt]{%
\raisebox{-#1}{\underline{\raisebox{#1}{\smash{\underline{#2}}}}}}
\newcommand{\Nt}{N_{t}}
\newcommand{\Ns}{N_{f}}
\newcommand{\Na}{N_{a}}
\newcommand{\Np}{N_P}

\newcommand{\tensor}[1]{\boldsymbol{\mathcal{#1}}}

\begin{document}
	
\title{PLAIN: Scalable Estimation Architecture for Integrated Sensing and Communication}

\author{Bashar Tahir, Philipp Svoboda, and Markus Rupp
    \thanks{Bashar Tahir and Philipp Svoboda are with Christian Doppler Laboratory for Digital Twin assisted AI for sustainable Radio Access Networks, Institute of Telecommunications, TU Wien, Austria. Markus Rupp is with the Institute of Telecommunications, TU Wien, Austria. The financial support by the Austrian Federal Ministry for Labour and Economy and the National Foundation for Research, Technology and Development, and the Christian Doppler Research Association is gratefully acknowledged.}
    \thanks{This article has optional downloadable code available through the link \href{https://github.com/bashar-tahir/plain}{https://github.com/bashar-tahir/plain}, provided by the authors.}
}
\maketitle
 \makeatletter
 \def\ps@IEEEtitlepagestyle{%
	   \def\@oddfoot{\mycopyrightnotice}%
	   \def\@oddhead{\hbox{}\@IEEEheaderstyle\leftmark\hfil\thepage}\relax
	   \def\@evenhead{\@IEEEheaderstyle\thepage\hfil\leftmark\hbox{}}\relax
	   \def\@evenfoot{}%
	 }
 \def\mycopyrightnotice{%
	   \begin{minipage}{\textwidth}
		   \centering \scriptsize
		   This work has been submitted to the IEEE for possible publication. Copyright may be transferred without notice, after which this version may no longer be accessible.
		   \end{minipage}
	 }
\begin{abstract}
Integrated sensing and communication (ISAC) is envisioned be to one of the paradigms upon which sixth-generation (6G) mobile networks will be built, extending localization and tracking capabilities, as well as giving birth to environment-aware wireless access. A key aspect of sensing integration is parameter estimation, which involves extracting information about the surrounding environment, such as the direction, distance, velocity, and orientation of various objects within. This is typically of a high-dimensional nature, which can lead to significant computational complexity, if performed jointly across multiple sensing dimensions, such as space, frequency, and time. Additionally, due to the incorporation of sensing on top of the data transmission, the time window available for sensing is likely to be very short, resulting in an estimation problem where only a single snapshot is accessible. In this article, we propose PLAIN, a tensor-based parameter estimation architecture that flexibly scales with multiple sensing dimensions and can handle high dimensionality, limited measurement time, and super-resolution requirements. It consists of three stages: a compression stage, where the high dimensional input is converted into lower dimensionality, without sacrificing resolution; a decoupled estimation stage, where the parameters across the different dimensions are estimated in parallel with low complexity; an input-based fusion stage, where the decoupled parameters are fused together to form a paired multidimensional estimate. We investigate the performance of the architecture for different configurations and compare its performance against practical sequential and joint estimation baselines, as well as theoretical bounds. Our results show that PLAIN, using tools from tensor algebra, subspace-based processing, and compressed sensing, can scale flexibly with dimensionality, while operating with low complexity and maintaining super-resolution capabilities.
\end{abstract}


\section{Introduction}
\IEEEPARstart{F}{or} the past couple of decades, the design of mobile networks has been shaped by the focus on increasing throughput, expanding connectivity, and reducing access latency. These aspects, or \acp{KPI}, have enabled the successful maturity of mobile networks into what they are today, becoming an everyday necessity of our lives. With the evolution of mobile networks towards the \ac{6G}, new use cases and applications necessitate the consideration of other \acp{KPI}, such as energy efficiency, awareness of the surrounding environment, and machine intelligence \cite{Dogra21,Lima21,Du20}. 

Our focus here will be on awareness, where \ac{RF} signals are used to sense the surrounding environment directly on the \ac{PHY}, essentially equipping the mobile network with radar functionalities. Such a combination of communication and radar systems, has led to much attention recently under the umbrella term of \ac{ISAC}, and it is now seen as one of the cornerstones for the evolution towards \ac{6G}. This paves the way for new applications, such as intelligent transportation and traffic management, smart factories, advanced localization and tracking, and vision and mapping through obstacles \cite{Wang22}. In addition to new applications, it also unlocks a new paradigm of mobile network optimization, where the knowledge of the surrounding environment is used to optimize the transmitted signals, such as environment-aware beamforming and obstacles-aware proactive transmissions \cite{Gonzalez24}.

Radar (or sensing) and communication systems have classically been considered separately with different requirements and targeting different deployments scenarios. The convergence of them together into a single system, i.e., \ac{ISAC}, can be categorized as being radar-centric, communication-centric, or a hybrid thereof \cite{Zhang22}. For mobile networks, the communication-centric approach is perhaps the more relevant one, as the focus remains primarily on data communications, and the sensing part would then come into play as a complementary addition, built upon the infrastructure provided by the mobile networks.

A fundamental task in sensing integration is parameter estimation, where information regarding the propagation environment is extracted, such as the direction, distance, velocity, and orientation of the different objects, including active devices, in the surrounding environment \cite{Zhang21}. This is generally a high-dimensional estimation problem, which can result in high computational complexity, if it is to be carried out jointly across the possible sensing dimensions, e.g., space, frequency, and time. Moreover, since the sensing is integrated on top of the data transmission and due to the possible rapid temporal variations, the time-window for sensing is likely to be very short, which results in an estimation problem with a single snapshot/measurement available. The high number of sensing dimensions, as well as the limited number of measurements, pose a significant challenge to many of the existing parameter estimation architectures. Therefore, it is important to investigate how the estimation task can be carried out in an efficient manner, while benefiting from the multidimensional structure and maintaining potential super-resolution capabilities.

\subsection{Related Work}
The problem of identifying objects from sensory readings and estimating their parameters spans multiple fields. This includes classical radar, as well as sonar, biomedical imaging, space interferometry, seismology, etc. \cite{Liu23}. Many solutions were proposed to tackle this problem across the different fields; therefore, there exists a rich set of literature on this topic, specifically under multidimensional harmonic retrieval.

For the application of integrating sensing capabilities into mobile networks, we focus here on categorizing the estimation algorithms into two classes. The first class belongs to algorithms that perform the estimation in a sequential manner, whether fully or partially \cite{Xu23,Islam22,Chen23,Tong23,Huang12,Keskin21}. The advantage of the sequential approach is that it can estimate the parameters one dimension at a time, and therefore it can maintain low estimation complexity when many sensing dimensions (e.g., azimuth, elevation, frequency, and time blocks) are involved. On the down-side, the sequential approach may ignore information that is present in the multidimensional structure of the input, which can be utilized to improve the estimation performance, especially with respect to the problem of association/pairing of the parameters belonging to the same object across the different dimensions. Another issue is related to error propagation, in the sense that if the estimation across one dimension fails, it can severely affect the estimation over the dimensions that follow.

The second class of algorithms are these that attempt to jointly estimate the parameters across the different dimensions \cite{Pesavento04,Zheng17,Wan18,Liu20,Sanson20,Masood21,Zhang23,Dehkordi23,Li24,Xiao24,Hu24}. The joint processing opens up the door for exploiting the multidimensional structure of the problem, and also naturally results in an automatic pairing of the parameters across the different dimensions. This, however, comes generally at a high computational cost. Moreover, many of the joint solutions are of an iterative nature that can suffer from convergence issues, or are restricted to a certain structure of the estimation problem. Of particular interest are sensing problems that have an inherent separable structure. In this case, the underlying system model can be described in tensor form, where each tensor mode would correspond to one of the estimation dimensions. This has sparked major interest, with many tensor-based algorithms being proposed to tackle the sensing problem \cite{Boyer08,Nion10,Zhou17,Costa19,Xu22,Zhang222,Zhao23,Chen24,Zhang24}. Although the majority focus on tensor decompositions with iterative solutions based on \ac{CPD}/PARAFAC, prior related work has been already carried out in extending the classical ESPRIT algorithm to the multidimensional tensor format, leading to the Tensor-ESPRIT algorithm \cite{Haardt08}. Not only does Tensor-ESPRIT provide competitive performance to recent iterative algorithms based on various tensor decompositions, but it also utilizes the Vandermonde structure present, similarly to other algorithms promoting that \cite{Xu22,Zhang212,Zhang24}.

Utilizing the tensor structure paves the ground for feasible implementations. However, issues with complexity, scalability, and maintaining super-resolution still remain open. In addition, the limited sensing time, resulting in a single snapshot available, puts further strain on many existing architectures.

\subsection{Contribution}
In this work, we address the aforementioned problems with the two classes of the sequential and joint approaches, and introduce an architecture that strikes a balance between complexity and performance, while providing high level of scalability with the dimensionality of the problem and the limited number of snapshots. Specifically, it goes as follows:
\begin{enumerate}
    \item Building on numerous works in the literature, we propose \ac{PLAIN}, a flexible and scalable estimation architecture applicable to an arbitrary number of sensing dimensions. The architecture utilizes the tensor structure of the estimation problem and breaks down the processing into three stages: first, a compression stage, where the original high dimensional input is brought down into lower dimensionality, suitable for low-complexity processing, while maintaining resolution. Second, a decoupled estimation stage, where the parameters across each of the sensing dimensions are estimated independently in parallel, thereby substantially reducing the estimation complexity. Third, a fusion stage, where the estimated parameters across the different dimensions are combined together to form a joint multidimensional estimate, achieving parameter pairing in the process.
    \item We investigate possible compression methods and discuss their implementation in tensor format. We discuss the implementation of the decoupled estimation stage for various one-dimensional algorithms, focusing on subspace methods, and consider practical estimation of the number of objects in the process. We then discuss suitable approaches for fusion and highlight an implementation that utilizes the sparsity of the problem using a tensor version of the \ac{OMP}, as well as a low-complexity approach based on one-shot \ac{LS} processing.
    \item We evaluate the performance of the proposed architecture via an \ac{OFDM}-based transmission using a setup suitable for a \ac{5G-NR} deployment. We compare the performance against a sequential baseline as well as the Tensor-ESPRIT algorithm. We also accompany our results with theoretical baselines obtained by the \ac{CRB}. Our results show that that the proposed \ac{PLAIN} architecture can provide super-resolution sensing capabilities within a single-snapshot setup, while flexibly scaling with dimensionality.	
\end{enumerate}

\subsection{Notation}
We denote vectors, matrices, and tensors by bold lower-case, bold upper-case, and bold calligraphic letters, e.g., $\mathbf{x}$, $\mathbf{X}$, and $\tensor{X}$, respectively, while sets are denotes by normal calligraphic letters, e.g., $\mathcal{I}$. The superscript symbols ${\mathbf{X}}^{\textsf{T}}$, ${\mathbf{X}}^{\textsf{H}}$, ${\mathbf{X}}^*$, and ${\mathbf{X}}^{\dagger}$ denotes the transpose, Hermitian transpose, complex conjugate, and pseudo-inverse of $\mathbf{X}$, respectively. The operations $\otimes$, $\diamond$, $\odot$, and $\circ$ denote the Kronecker product, Khatri-Rao product, Hadamard (element-wise) product, and tensor outer product, respectively. The operation ${|\tensor{X}|}^2$ applies the magnitude-squared operation element-wise to the tensor, i.e., $|\tensor{X} |^2 = \tensor{X} \odot \tensor{X}^*$. We denote by $\tensor{X}_{[m]}$ the matrix/slice obtained by the unfolding of tensor $\tensor{X}$ across its $m$-th dimension. Furthermore, we use $\times_m$ to denote the mode-$m$ product, i.e., $\tensor{X} \times_m \mathbf{U}$ corresponds to a transformation of the tensor $\tensor{X}$ across its $m$-th mode/dimension by matrix $\mathbf{U}$. We also use $\sqcup_m$ to denote the concatenation operation across the $m$-th mode. We use tensor mode and tensor dimension interchangeably, depending on the context. If we are discussing more abstract mathematical operations of the tensor itself, we use mode. Otherwise, if we are referring to a general description of the parameters to be estimated, then we use dimension. For example, when discussing angles over antennas, then we refer to it as the spatial dimension, rather than the spatial mode. 

We use the $\textrm{argsort}$ and $\textrm{argmax}$ functions in this work for multidimensional index searching. In order to avoid cluttering the equations with long lists of index variables, we omit the argument variables from the function, and the corresponding operations are implicitly understood as search operations over the corresponding multidimensional index-space.

\section{System Model for Sensing}
We consider an \ac{OFDM}-based transmission with the $\Ns$ subcarriers and $\Nt$ time symbols, with corresponding subcarrier- and time-spacing of $\Delta f$ and $\Delta t$, respectively. We consider, without loss of generality, the case of a single waveform illuminating the surrounding environment, and getting picked off by $\Na$ receive antennas of a \ac{ULA} with $\Delta a$ inter-antenna spacing. In order to support high resolution sensing, we consider operation on  the \ac{CSI}-level, where an estimate of the channel coefficients over space, frequency, and time is available at the receiver. Here, with time, we refer to the dimension across the different \ac{OFDM} symbols, not the time samples within one \ac{OFDM} symbol, i.e., slow-time. The channel can be estimated either using pilots, followed by interpolation across the time-frequency grid, or by utilizing the entire transmission block, if the data symbols are known as well. Consequently, the framework considered here is generally applicable to both monostatic, as well as bistatic configurations, and also for both the downlink and uplink. The validity of the model and the extension to multiple transmitted waveforms, is discussed later.

Modeling the incoming signal as an emergence from major objects, scatterers, or sources along $\Np$ paths, and under perfect channel estimation, the channel coefficient at receive antenna $r$, subcarrier $k$, and \ac{OFDM} symbol $s$ is given by
\begin{align}\label{eq:H}
    H_{r,k,s} = \sum_{p=1}^{\Np} \beta_p\, e^{j2\pi\big(\frac{1}{\lambda} \cos\theta_p\, r \Delta a \,-\, \tau_p k\Delta f \,+\, \upsilon_p s\Delta t\big)}\,,
\end{align}
where the parameters $\beta_p \in \mathbb{C}$, $\theta_p \in [0, \pi)$, $\tau_p \in \mathbb{R}^{+}$, and $\upsilon_p \in \mathbb{R}$ are the path gain, azimuth \ac{AOA}, delay, and Doppler shift of the $p$-th path, respectively. The path gain $\beta_p$ is complex-valued with random phase, while its magnitude is associated with the \ac{PL} and the object \ac{RCS}, i.e.,
\begin{align}
    |\beta_p| = \sqrt{\textrm{PL}_p\, \textrm{RCS}_p}\,.
\end{align}
In the monostatic case, where a reference plane is established by default, e.g., a \ac{BS} listening to the echos of its own downlink transmission, the parameters can be interpreted in an absolute manner. For example, the delay $\tau_p$ would correspond to the round-trip time between the \ac{BS} and object, and together with the angle $\theta_p$ enable the exact localization of that object. However, in some bistatic cases, e.g., uplink transmission from a \ac{UE} to a \ac{BS}, these parameters become relative to the propagation environment between the \ac{UE} and the \ac{BS}, and without an a priori established reference, such as a \ac{LOS} between the two, localization with respect to a global reference plane is challenging.

For the clarity of presentation, we stick with the estimation along the three dimensions of space (azimuth), frequency, and time (\ac{OFDM} symbols) when giving examples. However, the framework is applicable to any number of dimensions, e.g., we can extend \eqref{eq:H} with elevation as a fourth dimension, and to the case where multiple parameters are estimated per dimension (e.g., angle-distance estimation in a near-field setup). We focus mainly on the case where a single snapshot is available at the receiver, as it would be the more common case in the context of sensing in current mobile networks, but we also discuss how the formulation is extended to the multiple snapshots case.

\subsection{Problem Formulation}
The sensing task here is to estimate the parameters $\theta_p$, $\tau_p$, $\upsilon_p$, and also $\beta_p$. In order to efficiently tackle this multidimensional estimation problem, we first reformulate \eqref{eq:H} into a multilinear representation. Let
\begin{align}\label{eq:steeringvecs}
    \begin{split}
        \mathbf{a}_p &= [1, e^{j\frac{2\pi}{\lambda} \cos\theta_p\,\Delta a}, \dots, e^{j\frac{2\pi}{\lambda} \cos\theta_p\, (N_a-1)\Delta a}]^{\mathsf{T}}\,,\\
        \mathbf{d}_p &= [1, e^{-j2\pi \tau_p \Delta f}, \dots, e^{-j2\pi \tau_p (\Ns - 1)\Delta f}]^{\mathsf{T}}\,,\\
        \mathbf{v}_p &= [1, e^{j2\pi \upsilon_p \Delta t}, \dots, e^{j2\pi \upsilon_p (\Nt - 1)\Delta t}]^{\mathsf{T}}\,
    \end{split}
\end{align}
be the array, delay, and Doppler response (or steering) vectors of the $p$-th path, respectively. Stacking the channel coefficients in \eqref{eq:H} into a single vector $\mathbf{h} \in \mathbb{C}^{(\Na \Ns \Nt) \times 1}$, we obtain
\begin{align}\label{eq:h_vector}
	\mathbf{h} = \sum_{p=1}^{\Np} \beta_p \big( \mathbf{a}_p \otimes \mathbf{d}_p \otimes \mathbf{v}_p\big)\,,
\end{align}
where $\otimes$ denotes the Kronecker product. Let
\begin{align}
    \begin{split}
        \mathbf{A} &= [\mathbf{a}_1, \mathbf{a}_2, \dots, \mathbf{a}_{\Np}]\,,\\
        \mathbf{D} &= [\mathbf{d}_1, \mathbf{d}_2, \dots, \mathbf{d}_{\Np}]\,,\\
        \mathbf{V} &= [\mathbf{v}_1, \mathbf{v}_2, \dots, \mathbf{v}_{\Np}]\,,\\
        \mathbf{b} &= [\mathbf{\beta}_1, \mathbf{\beta}_2, \dots, \mathbf{\beta}_{\Np}]^{\mathsf{T}}\,,
    \end{split}
\end{align}
we can write \eqref{eq:h_vector} compactly as
\begin{align}\label{eq:h_vector_KR} 
    \mathbf{h} = \big(\mathbf{A} \diamond \mathbf{D} \diamond \mathbf{V}\big) \mathbf{b}\,,
\end{align}
where $\diamond$ denotes the column-wise Khatri-Rao product. 

The description using stacking (vectorization) clearly exposes the tensor structure of the problem, and already unlocks the door for simplified calculations. However, the treatment remains in an equivalent matrix-vector format with large dimensionality. To fully utilize the underlying tensor structure, we now switch to a multilinear tensor representation. Considering the channel coefficients in \eqref{eq:H} as a multidimensional variable over the three dimensions of space, frequency, and time, we can describe the channel directly as
\begin{align}\label{eq:H_outer}
    \tensor{H} = \sum_{p=1}^{\Np} \beta_p \big( \mathbf{a}_p \circ \mathbf{d}_p \circ \mathbf{v}_p\big)\,,
\end{align}
where $\tensor{H} \in \mathbb{C}^{\Na \times \Ns \times \Nt}$ is the third-order (space, frequency, and time) channel tensor and $\circ$ denotes the outer product. Viewing the matrices $\mathbf{A}, \mathbf{D},$ and $\mathbf{V}$ as transformations operating along the corresponding estimation dimensions, we can  further write \eqref{eq:H_outer} as
\begin{align}
    \tensor{H} = \tensor{B} \times_1 \mathbf{A} \times_2 \mathbf{D} \times_3 \mathbf{V}\,,
\end{align}
where $\tensor{B} = \text{sdiag}(\mathbf{b}) \in \mathbb{C}^{\Np \times \Np \times \Np}$ is the diagonal core tensor, with the path gains along its superdiagonal, and $\times_m$ denotes the mode-$m$ product, which corresponds to a transformation of the core tensor along the $m$-th mode/dimension. For an arbitrary number of dimensions $M$, this can be generally written as 
\begin{align}\label{eq:h_b_us}
    \tensor{H} = \tensor{B} \times_1 \mathbf{U}_1 \times_2 \mathbf{U}_2 \times_3 \dots \times_M \mathbf{U}_M\,,
\end{align}
where $\mathbf{U}_m \in \mathbb{C}^{N_m \times \Np}$ is the response/steering matrix along the $m$-th dimension, with $N_m$ being the response length. For our space-frequency-time example, we have $N_1 = \Na$, $N_2 = \Ns$, and $N_3 = \Nt$. If multiple snapshots are available, either naturally due to multiple measurement instances or artificially by post-processing techniques, then we extend our channel tensor with an $(M+1)$-th dimension corresponding to the measurements dimension. Let $S = N_{M+1}$ be the number of snapshots, and denote by $\tensor{H}^{(s)}$ the channel of the $s$-th snapshot, then our considered tensor for processing $\tensor{H} \in \mathbb{C}^{N_1 \times N_2 \times \dots \times N_{M} \times S}$ is given by
\begin{align}\label{eq:concat}
	\tensor{H} = \Big[\tensor{H}^{(1)} ~\sqcup_{M+1}~ \tensor{H}^{(2)} ~ \dots ~\sqcup_{M+1}~ \tensor{H}^{(S)}\Big]\,,
\end{align}
where $\sqcup_m$ denotes the concatenation operation across the $m$-th dimension, i.e., each $\tensor{H}^{(s)}$ becomes a slice across the $(M+1)$-th mode. In matrix-vector notation, this is similar to stacking column vectors horizontally to from a measurement matrix. For information regarding tensor algebra, refer to \cite{Cichocki15}.

\subsection{Validity of the Model}
We briefly discuss some aspects related to the assumptions made of the model in \eqref{eq:H}. The description given is clearly formulated for a narrowband system with time-invariant delays. For not highly-wideband systems and slowly-varying delays, the model can still be assumed to hold approximately, and therefore our formulation here is still applicable. The description also does not include possible timing- or frequency-offsets, as the sensing here is assumed to happen at the \ac{CSI}-level; some form of synchronization is already established.

Another aspect is with respect to multiple transmitted waveforms. The model holds automatically for a single waveform/layer transmission. This fits nicely for the uplink case, since the assumption of a user transmitting with a single stream holds in many cases. The model also holds directly for monostatic downlink setups during a discovery phase, where the \ac{BS} transmits the same signal through the transmit antennas \cite{Dehkordi23}, or when the sensing is carried out on a per-beam basis. Otherwise, if multiple waveforms are transmitted through the transmit antennas and picked up at a receiver for sensing, then data interference needs to be resolved. This can either be handled actively by trying to estimate the interference and perform a combination across the transmit spatial dimension as a pre-processing step, as in \cite{Xiao24}. Alternatively, we can rely on the fact that different layers use orthogonal pilots. This allows us to estimate the channel across the different transmit layers, and then use the estimated channel realizations across the transmit layers as multiple snapshots and concatenate them following \eqref{eq:concat}, or also combine them coherently as discussed.

Lastly, as our focus here is on the parameter estimation itself, we ignore aspects related to the full-duplex operation and how self-interference can be handled.

\section{PLAIN Architecture}
In this section, we introduce a flexible and scalable sensing architecture capable of utilizing the multidimensional structure and achieving super-resolution, while maintaining low operation complexity. We give it the name \acf{PLAIN}. 
\begin{figure*}[!t]
    \centering
    \includegraphics[width=0.87\textwidth]{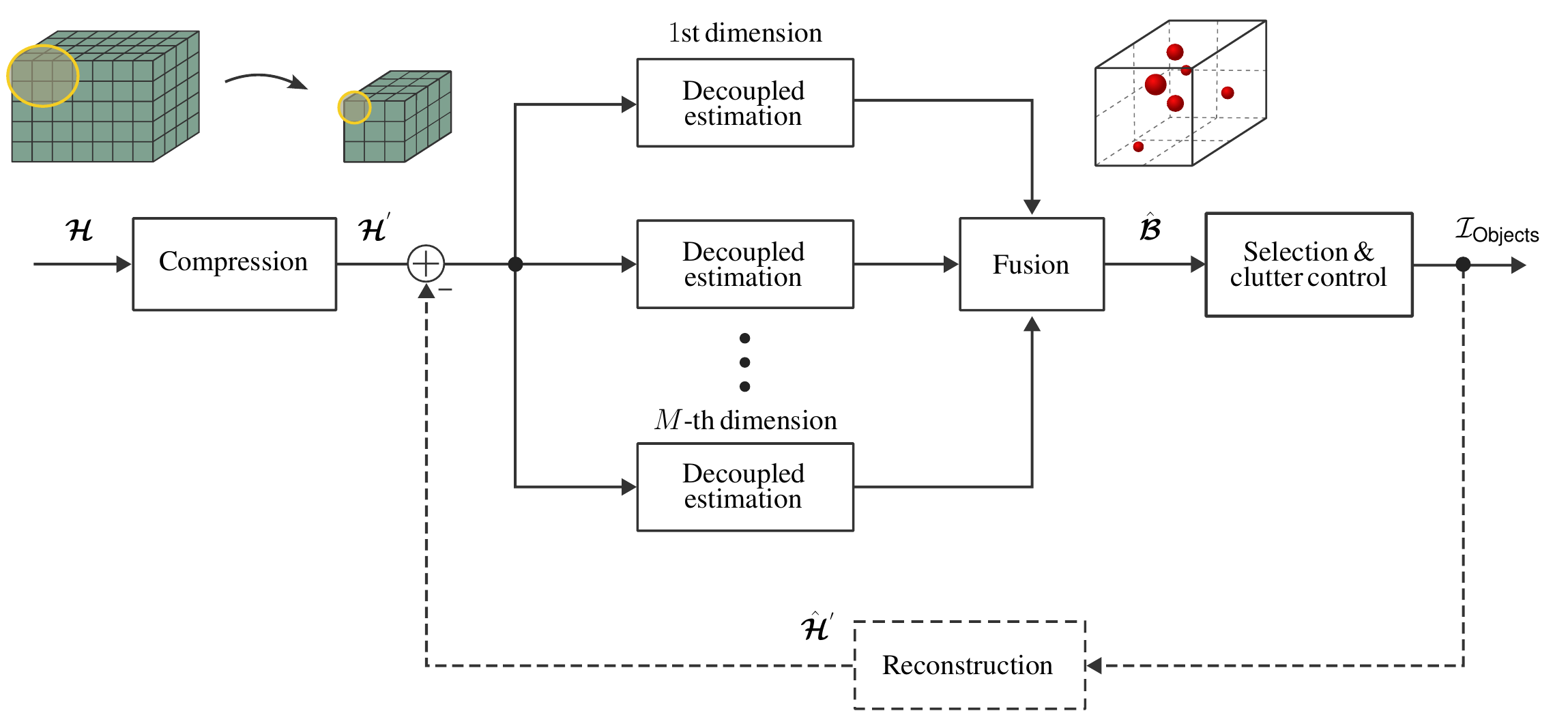}
    \captionsetup{justification=centering}
    \caption{Illustration of the proposed \ac{PLAIN} architecture with its main building blocks.}
    \label{fig:arch_1}
\end{figure*}
A sketch of the architecture is illustrated in Fig. \ref{fig:arch_1}, consisting of three stages: 
\begin{itemize}[leftmargin=1em,itemindent=0em]
    \item \textbf{Compression}, where the high-dimensional input $\tensor{H}$ is converted into a lower-dimensional representation $\tensor{H}^{'}$, by utilizing redundancy in the received signal based on prior information of the parameters, the underlying structure of the signal itself, or both. With the reduced dimensionality, low-complexity processing follows in the next stages.
    
    \item \textbf{Decoupled estimation}, where decoupled per-dimension estimation is carried out in parallel across the different separable dimensions. An interesting feature here is that an arbitrary combination of estimation algorithms can be applied across the different dimensions. For example, \ac{CS} can be used over the first estimation dimension, while a subspace-based technique is used over the second dimension, and a \ac{DFT}-based technique over the third one.
    
    \item \textbf{Input-based fusion and selection}, where the estimated parameters across the different dimensions together with the compressed input $\tensor{H}^{'}$ are fused to produce an estimate of the core tensor $\tensor{B}$, resulting in a joint multidimensional estimate and achieving automatic pairing of the parameters in the process. Finally, selection and clutter control are performed, where the active objects from the fused output are determined and possibly filtered down into a shortened list of major objects and active devices in the environment.	
\end{itemize}
Optionally, it is possible to run the process in an iterative manner: after each iteration, the contribution of the objects that are declared active is removed from the compressed input $\tensor{H}^{'}$, and another around of estimation is performed, mimicking a CLEAN-like procedure \cite{Tsao88}. This has the potential of detecting weak objects that are dominated by strong ones; however, this is likely to increase the false alarm rate. We illustrate that with the dashed part in the figure. 

Next, we go through each of those stages and discuss candidate implementations and potential issues.

\subsection{Compression}\label{section:compression}
Although high dimensionality is generally a requirement for high-resolution capabilities, it can be a curse when it comes to implementation and processing complexity. Consider a system with $\Na = 16$ antennas, $\Ns = 180$ subcarriers, and $\Nt = 560$ symbols. For $\Delta f = 60$\,kHz, this would correspond to a bandwidth of $10.8$\,MHz and a transmission (sensing) time of $10$\,ms, including the \ac{CP}. The compound dimensionality of this problem, i.e., of \eqref{eq:h_vector_KR}, is $\Na \Ns \Nt = 1612800$. Decoupling the estimation task into separate dimensions can help already in reducing the complexity; however, it can still be too high. For example, if we would like to apply a subspace-based method for high-resolution delay estimation, then this would generally require an eigendecomposition of a size $180 \times 180$ autocorrelation matrix.

Fortunately, in many cases, certain deployment scenarios, target requirements, and structure can introduce redundancy in the received signal, which can help in reducing the dimensionality. Continuing with the previous example, if we operate in the \ac{mmWave} band, such as $26\,$GHz, then due to the high path loss, and in addition to the limitation of the \ac{CP}, we will likely only be able to resolve close objects. In this case, we can limit our round-trip distance of interest to, e.g., $1$\,km, which corresponds to a delay of around $3.33\,\mu$s. In addition, since the system is envisioned to operate in urban areas, relative velocities beyond $\pm 80\,$km/s are rare. With these practical constraints, we can downsample $\eqref{eq:H}$ in the frequency and time symbol dimensions, without losing the ability of recovering the objects, as long as the resultant sampling rates across the dimensions do not violate the recovery requirements of $1$\,km distance and $\pm 80\,$km/h velocity, respectively. By straightforward verification, one can downsample the frequency dimension by a factor of $4$ and the time dimension by factor of $14$, without introducing aliasing/ambiguity. This results in the downsampled tensor $\tensor{H}^{'} \in \mathbb{C}^{\Na^{'} \times \Ns^{'} \times \Nt^{'}}$, with new dimensions $\Na^{'} = 16$, $\Ns^{'} = 45$, and $\Nt^{'} = 40$, where we left the spatial dimension intact. The compound dimensionality is then $\Na^{'} \Ns^{'} \Nt^{'} = 28800$, which is a reduction of $98\%$ compared to the original input.

In the following, we describe four possible compression/downsampling techniques. We start with these that do not produce additional virtual snapshots in the process, and then consider the other case with virtual snapshots, including the classical smoothing technique. These are illustrated in Fig. \ref{fig:compression} for a \ac{1D} tensor.
\subsubsection{Decimation}
The simplest method for compression is to decimate the input by certain factors across the different dimensions: for dimension $m$, keep every $\Delta_m$ sample, while throwing away the rest, where $\Delta_m$ is the compression factor along the $m$-th dimension. This results in the reduced dimensionality $N_m^{'} = N_m / \Delta_m\,.$

\begin{figure}[!t]
    \centering
    \subfloat[Decimation.\label{fig:compression:a}]{\includegraphics[width=0.225\textwidth]{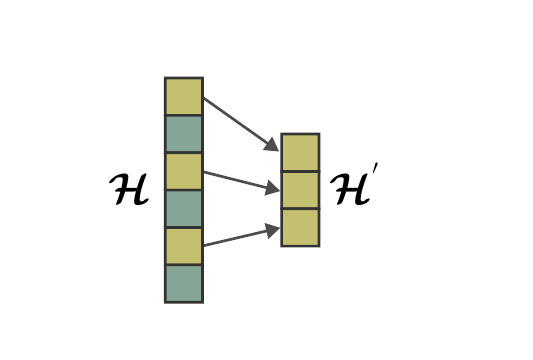}}
    \subfloat[Averaging.]{\includegraphics[width=0.225\textwidth]{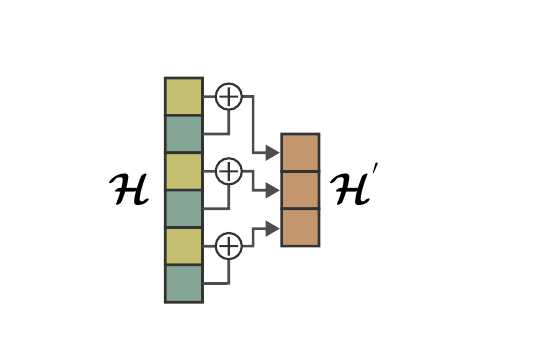}}
    \\
    \subfloat[Decimation; multiple snapshots.\label{fig:compression:c}]{\includegraphics[width=0.225\textwidth]{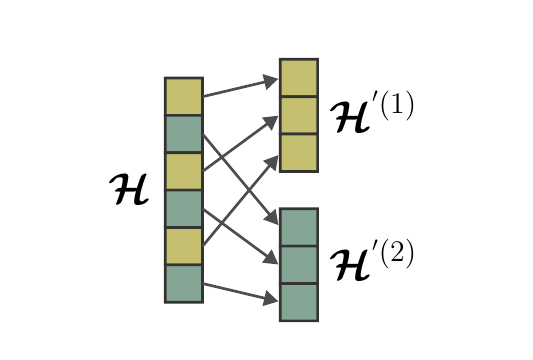}}
    \hspace{1em}\subfloat[Smoothing; multiple snapshots.]{\includegraphics[width=0.225\textwidth]{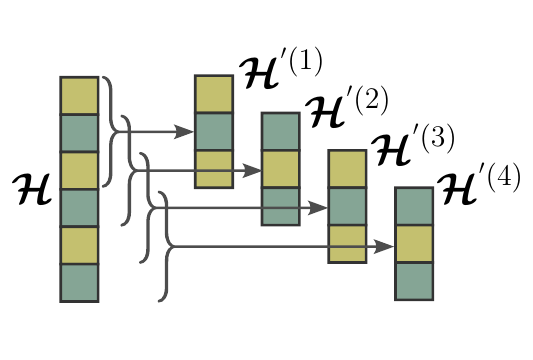}}
    \caption{Illustration of the different compression methods for a \ac{1D} tensor.}
    \label{fig:compression}
\end{figure}

We stick to the multilinear format and write the decimation operation in tensor description directly. Let $\mathbf{e}_{i,m} \in \mathbb{R}^{N_m}$ be a column vector with a one at index $i$ and zeros otherwise. We can construct the selection matrix for the decimation operation over the $m$-th dimension $\mathbf{J}^{\mathsf{d}}_{m} \in \mathbb{R}^{N^{'}_m \times N_m}$ as
\begin{align}\label{eq:decimat}
    \mathbf{J}^{\mathsf{d}}_{m} = [\mathbf{e}_{0,m},\,\mathbf{e}_{\Delta_m,m},\,\dots, \,\mathbf{e}_{(N_m^{'}-1)\Delta_m,m}]^{\mathsf{T}}\,.
\end{align}
For $N_m^{'}$ even, $\mathbf{e}_{(N_m^{'}-1)\Delta_m,m}$ would be an all-zero vector, corresponding to throwing away the last element, as depicted in Fig. \ref{fig:compression:a}. For $\Delta_m = 1$, we have $N_m^{'} = N_m$ and therefore $\mathbf{J}^{\mathsf{d}}_{m}$ would simply be the identity matrix. The decimation operation can be described in tensor notation as
\begin{align}
    \tensor{H}^{'} = \tensor{H} \times_1 \mathbf{J}^{\mathsf{d}}_{1} \times_2 \mathbf{J}^{\mathsf{d}}_{2} \times_3 \dots \times_M \mathbf{J}^{\mathsf{d}}_{M}\,.
\end{align}
The compressed model is then represented with the downsampled response matrices
\begin{align}\label{eq:U_de}
    \mathbf{U}^{'}_m = \mathbf{J}^{\mathsf{d}}_{m}\mathbf{U}_m\,,
\end{align} 
which are used later during estimation and fusion. Note that with decimation, resolvability of the objects is maintained; only the estimation range of the parameter is limited. This is beneficial compared to classical smoothing, where aperture/resolvability of closely-spaced objects is sacrificed. For $N_m$ even, this statement holds approximately, since decimation would drop the last element of the response vector.

The choice of $\Delta_m$ should avoid producing aliasing, as well as making sure that the output length is an integer. Generally speaking, for a scenario with $\Np$ objects having sufficiently spaced-apart parameters, e.g., sufficiently different $\theta_p$, $\tau_p$, and $\upsilon_p$, it is required that $\min_m N_m^{'} \geq \Np$ to guarantee simultaneous resolvability across each dimension. Otherwise, recovery of $\Np$ components from each dimension would not be possible. If the objects share exact or very similar values for the parameters, than this condition can be relaxed.

\subsubsection{Averaging}
Instead of throwing away samples, we can make the assumption that the signal does not vary substantially across neighboring samples, and therefore every $\Delta_m$ samples in the $m$-th dimension can be averaged together to form a single sample in the compressed tensor $\tensor{H}^{'}$. Under the perfect assumptions of the signal component being constant within the averaging interval and having statistically independent noise samples, averaging improves the \ac{SNR} across the $m$-th dimension by factor of $\Delta_m$. 

Let $\mathbf{e}_{i,m}\in \mathbb{R}^{N_m}$ be a column vector with ones at indices $i$ to $(i + \Delta_m -1)$, and zeros otherwise. The selection matrix for the averaging operation $\mathbf{J}^{\mathsf{a}}_{m} \in \mathbb{R}^{N^{'}_m \times N_m}$ is built as
\begin{align}
    \mathbf{J}^{\mathsf{a}}_{m} = \frac{1}{\Delta_m} [\mathbf{e}_{0,m},\,\mathbf{e}_{\Delta_m,m},\, \dots, \,\mathbf{e}_{(N_m^{'}-1)\Delta_m,m}]^{\mathsf{T}}\,,
\end{align}
with the corresponding multidimensional-averaged tensor
\begin{align}
    \tensor{H}^{'} = \tensor{H} \times_1 \mathbf{J}^{\mathsf{a}}_{1} \times_2 \mathbf{J}^{\mathsf{a}}_{2} \times_3 \dots \times_M \mathbf{J}^{\mathsf{a}}_{M}\,.
\end{align}
Note that the decimated responses matrices in \eqref{eq:U_de} would also be used here for the estimation. In other words, averaging assumes the same input-output relationship as decimation, while resulting in an improved \ac{SNR}. This holds if the response varies slowly across neighboring samples, which is the case for, e.g., the Doppler response. However, for the spatial array response, the phase shifts between two neighboring elements can be substantial, which might make averaging across the spatial domain problematic. In practice though, we would avoid compression along the spatial domain, since this would limit the angular resolvability (which is important) and also the number of antennas is usually not very large.

\subsubsection{Decimation with Virtual Snapshots}
The previous two approaches reduce the dimensionality by reducing the number of samples across each dimension. Alternatively, we can utilize the redundancy in the received signal by forming virtual snapshots, instead of throwing away samples or averaging them, as depicted in Fig. 
\ref{fig:compression:c}. Whether the creation of additional snapshots is beneficial or not, depends on the employed estimation algorithms afterwards and how they deal with the presence of multiple snapshots.

Let $\mathbf{J}^{\mathsf{d}}_{m,i}$ be the selection matrix in \eqref{eq:decimat} with its columns circularly shifted by $i$ positions, i.e., $\mathbf{J}^{\mathsf{d}}_{m,i} = \mathbf{J}^{\mathsf{d}}_{m}\mathbf{C}_i$, where $\mathbf{C}_i$ is the permutation matrix for the circular shift. The compressed tensor is then formed by the concatenation of all possible shifted decimations over the $(M+1)$-th dimension, i.e., 
\begin{align}\label{eq:decishifts}
    \tensor{H}^{'} = \Big[\mathlarger{\sqcup}_{M+1} (\tensor{H} \times_1 \mathbf{J}^{\mathsf{d}}_{1,i_1} \times_2 \mathbf{J}^{\mathsf{d}}_{2,i_2} \dots \times_M \mathbf{J}^{\mathsf{d}}_{M,i_M})  \Big]\,,
\end{align}
for all possible shifts $i_1, i_2, \dots, i_M$ across the $M$ dimensions.

\subsubsection{Smoothing}
This is a classical approach that utilizes the rotational invariance structure of \eqref{eq:steeringvecs}, in which a subvector of these responses can be written as a shifted version of the other subvectors. This transforms the original high dimensional problem into a lower dimensional problem, while allowing the generation of virtual snapshots in the process. The disadvantage of smoothing is that the subvectors span a smaller aperture than the original response vector, and therefore can affect the resolution capabilities of the estimation architectures afterwards. Additionally, it is generally only applicable to response vectors that have the property of rotational invariance. For arbitrary response vectors, such a construction is not valid.

We build the smoothing selection matrix as $\mathbf{J}^{\mathsf{s}}_{m} = [\mathbf{I}_{N_m^{'}}~\mathbf{0}_{(N_m -N_m^{'})}]^{}$, where $\mathbf{I}_{n}$ and $\mathbf{0}_{n}$ are the identity and zero matrix of dimension $n\times n$, respectively, which selects the first $N_m^{'}$ elements of the input. The selection over the entire span of the input is then achieved by a circular shift of this matrix, i.e., we construct the shifted smoothing matrix $\mathbf{J}^{\mathsf{s}}_{m, i}$ in a similar manner as in \eqref{eq:decishifts}. The smoothed output tensor with multiple snapshots is finally constructed as
\begin{align}\label{eq:smoothing}
    \tensor{H}^{'} = \Big[\mathlarger{\sqcup}_{M+1} (\tensor{H} \times_1 \mathbf{J}^{\mathsf{s}}_{1,i_1} \times_2 \mathbf{J}^{\mathsf{s}}_{2,i_2} \dots \times_M \mathbf{J}^{\mathsf{s}}_{M,i_M})  \Big]\,,
\end{align}
for all possible shifts $i_1, i_2, \dots, i_M$ across the $M$ dimensions, with the effective response matrices given by
\begin{align}\label{eq:U_de_s}
    \mathbf{U}^{'}_m = \mathbf{J}^{\mathsf{s}}_{m}\mathbf{U}_m\,.
\end{align} 
The number of virtual snapshots created in the process can be very big for large compression factors. In order keep the complexity low, we consider a limited number of shifts, e.g., we only form $S = 100$ snapshots. Which shifts to keep can either be done in a random manner, or by choosing uniformly-sampled shifts across the different dimensions.

\subsection{Decoupled Estimation}
For the parameter estimation stage, three aspects are of interest: first, we try to avoid joint multidimensional operations. This is achieved already by utilizing the inherit tensor structure, allowing for a decoupled estimation across the different tensor dimensions. Second, we attempt to reduce the estimation complexity by reducing the dimensionally of the individual estimation tasks. This is achieved by compression in the previous stage. Third, we maintain a general description of the input and output of each decoupled estimation. This allows us to combine different estimation algorithms across the different dimensions. For example, we could use compressed sensing for angle estimation, while using a subspace-based method for delay, and a classical \ac{DFT} estimation for Doppler.

In order to support high-resolution sensing, we focus on grid-less parameter estimation. We demonstrate here how the estimation stage can be realized by running \ac{1D} root-MUSIC instances across the different dimensions \cite{Boyer08}. 

\subsubsection{Subspace-Based Estimation}
Consider the the $m$-th mode unfolding of the tensor $\tensor{H}^{'}$, $\tensor{H}^{'}_{[m]} \in \mathbb{C}^{N^{'}_m \times (N^{'}_1\dots N^{'}_{m-1}N^{'}_{m+1}\dots N^{'}_{M+1})}$, where $N^{'}_{M+1} = 1$ in the case of a single snapshot. We can estimate the $m$-th mode autocorrelation matrix $\mathbf{R}_m \in \mathbb{C}^{N^{'}_m \times N^{'}_m}$  as
\begin{align}
    \mathbf{R}_m = \frac{N^{'}_m}{\prod_i^{M+1}N^{'}_i}  \tensor{H}^{'}_{[m]} \tensor{H}^{'\textsf{H}}_{[m]}\,.
\end{align}
In other words, all other modes are combined into an extended snapshots dimension for the estimation of parameters along the $m$-th tensor dimension/mode. With access to the autocorrelation matrix, subspace methods can be readily used, which utilizes the separation of the input space into a signal-plus-noise and noise-only subspaces. Assuming $\Np$ objects are present, the noise subspace $\mathbf{Q}_{m,\textsf{noise}}$ is given by the collection of eigenvectors of $\mathbf{R}_m$ corresponding to the smallest $N^{'}_m - \Np$ eigenvalues. Let $\phi$ be the current parameter of interest, with a corresponding response vector $\mathbf{u}^{'}_m(\phi)$, the estimation problem is formulated as root-finding of the MUSIC spectrum, i.e.,
\begin{align}\label{eq:rootmusic}
    {\lVert \mathbf{Q}^{\textsf{H}}_{m,\textsf{noise}} \mathbf{u}^{'}_m(\phi) \rVert}^2 = 0\,.
\end{align}
The parameters $\phi$ solving this, are then declared to be active. Note that here the downsized response vector is used, i.e., after applying \eqref{eq:U_de} in the case of decimation and averaging, or \eqref{eq:U_de_s} in the case of smoothing. The parameter $\phi$ is arbitrary, e.g., it could correspond to the \ac{AOA}, delay, or Doppler. For the particular structure of the response/steering vectors in \eqref{eq:steeringvecs}, the calculation is simplified, as it can be viewed as finding the closest $\Np$ zeros to the unit-circle on the complex $z$-plane.
To further improve the estimation of the subspaces, \ac{FBA} can be used \cite{Linebarger94,Haardt08}. The autocorrelation matrix under \ac{FBA} is constructed via
\begin{align}
    \mathbf{R}_{m, \textsf{FBA}} = \frac{1}{2}\Big(\mathbf{R}_m + \mathbf{J}_{\textsf{FBA}} \mathbf{R}^*_m \mathbf{J}_{\textsf{FBA}} \Big)\,,
\end{align}
where $\mathbf{J}_{\textsf{FBA}}$ is the anti-identity matrix with ones on its anti-diagonal, while being zero elsewhere, and ${}^{*}$ denotes the complex conjugation. Matrix $\mathbf{R}_{m, \textsf{FBA}}$ is then used in place of the original $\mathbf{R}_{m}$ and the processing continues as before.

\subsubsection{Model-Order Determination}
The discussion in the previous part assumed that the number of objects $\Np$ is known a priori. However, in practice, we do not have access to that. In the context of subspace-based estimation, this plays a crucial role in determining the noise subspace dimensionality, as well as in knowing how many roots to consider from \eqref{eq:rootmusic}. This holds for other estimation algorithms as well. For example, if a compressed sensing algorithm is used instead, then the sparsity of the solution needs to be known. One can always use threshold-based techniques; however, the selection of the threshold then becomes the problem. An important aspect here is that the different dimensions can have different number of active components. For example, consider the scenario where three objects lie along the same direction to the \ac{BS}, but with different resolvable distances. Across the spatial dimension only one angle will be present; however, in the delay dimension, there will three distinct peaks.
 
To address that, we estimate the number of active components separately across the different tensor dimensions. Continuing with subspace-based estimation, since we already have access to the eigenvalues of the autocorrelation matrix for the $m$-th dimension, then information-theoretic criteria can be applied, such as the \ac{AIC} \cite{Stoica04}. If \ac{FBA} is used, then this would require modifying the \ac{AIC} calculation to account for it. However, in our experiments, using the original number of snapshots, i.e., $N^{'}_m\,/\,{\prod_i^{M+1}N^{'}_i}$ yielded better results and consistency than the formulation that accounts for \ac{FBA}. We denote the estimated number of components using \ac{AIC} over the $m$-th dimension as $\hat{N}_{P,m}$.

\subsubsection{Response Matrices Reconstruction}
Lastly, the response matrices are reconstructed. Let $\hat{\phi}_{m,1}, \hat{\phi}_{m,2}, \dots, \hat{\phi}_{m,\hat{N}_{P,m}}$ be the parameters estimated via \eqref{eq:rootmusic}. The reconstructed response matrix over the $m$-th dimension is then given by
\begin{align}\label{eq:resp_esi}
    \hat{\mathbf{U}}^{'}_m = \Big[\mathbf{u}^{'}_m(\hat{\phi}_{m,1}),\, \mathbf{u}^{'}_m(\hat{\phi}_{m,2}), \,\dots,\, \mathbf{u}^{'}_m(\hat{\phi}_{m,\hat{N}_{P,m}})\Big]\,.
\end{align}

\subsection{Input-Based Fusion and Selection}
Performing the estimation in a decoupled manner substantially reduces the estimation complexity. But, as the parameters are not estimated jointly, it is not possible to directly link the parameters of the same object across the different dimensions to each other. In order to tackle this, we fuse the parameters across the different dimensions by estimating the core tensor linking them. Using the estimated response matrices in \eqref{eq:resp_esi}, we can write snapshot $s$ of $\tensor{H}^{'}$ as 
\begin{align}
    \tensor{H}^{'(s)} = \tensor{B}^{(s)} \times_1 \hat{\mathbf{U}}^{'}_1 \times_2 \hat{\mathbf{U}}^{'}_2 \times_3 \dots \times_M \hat{\mathbf{U}}^{'}_M\,.
\end{align}
where $\tensor{B}^{(s)} \in \mathbb{C}^{\hat{N}_{P,1}\times \hat{N}_{P,2} \times \dots \times \hat{N}_{P,M}}$ is the core tensor of the $s$-th snapshot. By recovering $\tensor{B}^{(s)}$, we can establish an association between the parameters of the different response matrices. More specifically, the indices at which $\tensor{B}^{(s)}$ have large values, correspond to an active connection between the responses, and therefore indicate that these responses are linked. At the same time, the value of $\tensor{B}^{(s)}$ at these indices gives us an estimate of the corresponding path gains $\beta_p$. This allows to simultaneously pair the parameters of a certain object and estimate its corresponding path gain. 

In the following, we investigate two approaches for the recovery of the core tensor: one follows the \ac{LS} principle, while the second attempts to utilize the sparse structure of the problem via \ac{OMP}.

\subsubsection{Tensor-LS Fusion}
A straightforward approach to reconstructing the core tensor, is to attempt to reverse the effect of each transformation across the individual modes in the \ac{LS} sense. Denoting by ${}^\dag$ the pseudo-inverse operation, the core tensor of the $s$-th snapshot can be recovered as follows
\begin{align}\label{eq:ls_s}
    \hat{\tensor{B}}^{(s)}_{\textsf{LS}} =  \tensor{H}^{'(s)} \times_1 \hat{\mathbf{U}}^{'\dag}_1 \times_2 \hat{\mathbf{U}}^{'\dag}_2 \times_3 \dots \times_M \hat{\mathbf{U}}^{'\dag}_M\,.
\end{align}
Alternatively, instead of looking at each slice $s$ separately, we can also rewrite the previous calculation directly in terms of the multi-snapshot input $\tensor{H}^{'}$ as
\begin{align}
    \hat{\tensor{B}}_{\textsf{LS}} =  \tensor{H}^{'} \times_1 \hat{\mathbf{U}}^{'\dag}_1 \times_2 \hat{\mathbf{U}}^{'\dag}_2 \times_3 \dots \times_M \hat{\mathbf{U}}^{'\dag}_M\,,
\end{align}
where $\hat{\tensor{B}}_{\textsf{LS}} \in \mathbb{C}^{\hat{N}_{P,1}\times \hat{N}_{P,2} \times \dots \times \hat{N}_{P,M}  \times S}$ is the multi-snapshot core tensor. 
Under perfect recovery, the estimated core tensor is non-zero only at the positions where the response vectors are linked, i.e., a shuffled version of the original superdiagonal core in \eqref{eq:h_b_us}, with the non-zero elements corresponding to the path gains. The actual \ac{LS} estimate, however, will likely produce a non-sparse solution, and therefore we need to look for the strongest coefficients. Assuming $S$ snapshots are available, then we consider sorting according to the mean path-power across the snapshots, i.e.,
\begin{align}\label{eq:pathgains}
    \mathcal{I} = \textrm{argsort}~ \frac{1}{S}\sum_{s=1}^{S}{\big|\hat{\tensor{B}}^{(s)}_{\textsf{LS}}\big|^2}\,,
\end{align}
where ${|.|}^2$ applies the magnitude-squared operation element-wise to the tensor, i.e., $|\tensor{X} |^2 = \tensor{X} \odot \tensor{X}^*$ with $\odot$ being the Hadamard product, and the $\textrm{argsort}$ function returns the indices of the tensor elements sorted in descending order. More specially, the returned set $\mathcal{I}$ is defined as
\begin{align}\label{eq:pathgains2}
    \mathcal{I} = \{\mathbf{i}_1, \mathbf{i}_2, \dots, \mathbf{i}_{\hat{N}_{P,1}\hat{N}_{P,2}\dots\hat{N}_{P,M}}\}\,,
\end{align}
with vector $\mathbf{i}_p = [i_{p,1}, i_{p,2}, \dots, i_{p,M}]^{\textsf{T}}$ containing the corresponding indices across the $M$ modes of the core tensor. For example, $\mathbf{i}_5 = [3,2,4]^{\textsf{T}}$, indicates that the $5$-th strongest path is given by the third, second, and fourth columns of $\hat{\mathbf{U}}^{'}_1$, $\hat{\mathbf{U}}^{'}_2$, and $\hat{\mathbf{U}}^{'}_3$, respectively. Therefore, the corresponding parameters for the $5$-th object are given by $\hat{\phi}_{1,3}$, $\hat{\phi}_{2,2}$, $\hat{\phi}_{3,4}$, respectively. For the path gains $\beta_p$, since the phase varies from one snapshot to another, each $\hat{\tensor{B}}^{(s)}_{\textsf{LS}}$ will have a different phase estimate. As for the magnitude $|\beta_p|$, and assuming the parameters remain static within the sensing time, then an estimate can be obtained directly from mean tensor in \eqref{eq:pathgains}.

The question that remains is how many of the total indices are active objects of interest, and how many correspond to noise or clutter? A direct approach is to define a threshold for the mean path-power in \eqref{eq:pathgains}. All elements exceeding that threshold are then declared to be active. Alternatively, since we already have access to the number of active components over each tensor mode from the previous estimation stage, i.e., $\hat{N}_{P,m}$, then we can use this information to decide on the number of objects. We determine the number of active objects $\hat{N}_{P}$ as the maximum number of components detected over all of the tensor dimensions/modes, i.e.,
\begin{align}\label{eq:npmax}
    \hat{N}_{P} = \max_m\, \hat{N}_{P,m}\,.
\end{align}
For example, if three components are observed in the angular dimension, while only one is visible over the Doppler, e.g, objects being static, then we still declare that three objects are present. This is a simple approach that utilizes the multidimensional structure. To this end, we derive the final set of active objects as
\begin{align}
	\mathcal{I}_{\textsf{Objects}} = \mathcal{I}_{1:\hat{N}_{P}}\,,
\end{align}
i.e., the first $\hat{N}_{P}$ objects are selected from \eqref{eq:pathgains2}.

\subsubsection{Tensor-OMP Fusion}
The \ac{LS}-based approach ignores the sparse structure of the core tensor; this can be utilized to improve the probability of correctly associating the parameters across the different dimensions. Here, we view the fusion procedure as a support-recovery problem, where active columns of the response matrices are identified. As we aim for low-complexity, we utilize a tensor variant of \ac{OMP} \cite{Costa19}.

Similar to  classical \ac{OMP}, we attempt to iteratively identify the active components, starting with an empty set $\mathcal{I} = \{\}$. We begin with a matching step, where a correlation-based estimation is used to identify the strongest occurring connection in the core tensor. Assuming $S$ snapshots are available, the strongest active index $\mathbf{i}_p$ is identified as
\begin{align}
    \mathbf{i}_p = \textrm{argmax}~ \sum_{s=1}^{S}{\big| \tensor{H}^{'(s)} \times_1 \hat{\mathbf{U}}^{'\textsf{H}}_1 \times_2 \dots \times_M \hat{\mathbf{U}}^{'\textsf{H}}_M \big|^2}\,,
\end{align}
where $\textrm{argmax}$ returns the multidimensional index of the largest value from the sum tensor. The set $\mathcal{I}$ is then augmented with the detected index vector, i.e., $\mathcal{I} = \mathcal{I} \cup \{\mathbf{i}_p\}$, with the same structure as in \eqref{eq:pathgains2}. Let $\mathcal{I}_m$ denotes the indices set of the detected responses so far across the $m$-th mode, an approximation of the core tensor is then obtained as
\begin{align}
    \hat{\tensor{B}}_{\textsf{OMP}} =  \tensor{H}^{'} \times_1 \hat{\mathbf{U}}^{'\dag}_{1, \mathcal{I}_1}  \times_2 \hat{\mathbf{U}}^{'\dag}_{2, \mathcal{I}_2} \times_3 \dots \times_M \hat{\mathbf{U}}^{'\dag}_{M, \mathcal{I}_M}\,,
\end{align}
where $\hat{\mathbf{U}}^{}_{m, \mathcal{I}_m}$ is the submatrix formed by columns of $\hat{\mathbf{U}}^{}_{m}$ corresponding to unique indices in $\mathcal{I}_m$, i.e., no repeated columns. A low-rank approximation of $\tensor{H}^{'}$ is constructed as
\begin{align}
    \hat{\tensor{H}}\vphantom{\tensor{H}}^{'} =  \hat{\tensor{B}}_{\textsf{OMP}} \times_1 \hat{\mathbf{U}}^{'}_{1, \mathcal{I}_1}  \times_2 \hat{\mathbf{U}}^{'}_{2, \mathcal{I}_2} \times_3 \dots \times_M \hat{\mathbf{U}}^{'}_{M, \mathcal{I}_M}\,.
\end{align}
Finally, a residual is constructed through 
\begin{align}
	\tensor{R} = \tensor{H}^{'} - \hat{\tensor{H}}\vphantom{\tensor{H}}^{'}\,.
\end{align}
The process is repeated again in order to detect the remaining active objects, with the residual being used in the matching step instead of the original $\tensor{H}^{'}$. 
The maximum number of iterations determines the number of objects that are declared active. Here, we follow a similar approach as in the \ac{LS}-based fusion and set it according to \eqref{eq:npmax}.

The entries of $\hat{\tensor{B}}_{\textsf{OMP}}$ from the last iteration over the active set $\mathcal{I}_{\textsf{Objects}}$ provide us with an estimate of the complex path gains across the different snapshots. To obtain an estimate of the magnitude $|\beta_p|$, a mean estimate can be calculated, in a fashion similar to \eqref{eq:pathgains}.

\section{Performance Evaluation}
We evaluate the \ac{PLAIN} architecture using different configurations and compare it against two schemes, representing the two categories of sequential per-dimension estimation and tensor-based multidimensional estimation. The results are accompanied by the theoretically possible estimation performance established via the approximate\footnote{Greatly simplifies the calculations with the tensor format. However, it can become loose under certain configurations. See Section 8.4.1.1 of \cite{VanTrees04}.} deterministic \ac{CRB} \cite{VanTrees04}, applied to the original input $\tensor{H}$ corrupted by noise. 

\subsection{Setup}
We consider an \ac{OFDM}-based system operating at $26\,$GHz with the parameters listed in Table \ref{Table:t1}. As our focus is on sensing within a short transmission period, we consider the scenario where the received signal corresponds to a single snapshot. Therefore, our discussion of snapshots in the following is related only to those generated virtually.

\begin{table}[t]
	\begin{center}
		\small
		\caption{Base scenario parameters.}
		\label{Table:t1} 
		\begin{tabularx}{1\linewidth}{l |c}
			\hline
			\textbf{Parameter} & \textbf{Value} \\ \specialrule{1pt}{0pt}{1pt}
			Waveform & \ac{OFDM}  \\
			Center frequency & $26\,$GHz  \\
			\# antennas $\Na$ & 16 \\
			\# subcarriers $\Ns$ & 180 \\
			\# \ac{OFDM} symbols $\Nt$ & 560 \\
			Subcarrier spacing $\Delta f$ & $60\,$kHz  \\
			Bandwidth and sensing time  & $10.8\,$MHz and $10\,$ms  \\
			Path loss model and \ac{RCS} &  3GPP UMa LOS \cite{3GPP_TR_38901} and 1 \\
			Angles range & $[30, 150]^{\circ}$ (sector) \\
			Distances range &  $[50, 400]\,$m \\
			Velocities range &  $[0, 25]\,$km/h \\
			\ac{PLAIN} compression factors & $\Delta_1 = 1$, $\Delta_2= 4$, $\Delta_3= 14$\\
			\ac{PLAIN} decoupled estimation & Root-MUSIC with \ac{FBA} and \ac{AIC}\\
			\hline
		\end{tabularx}
	\end{center}
	\vspace{-0.5em}
\end{table}

In our base setup, we follow a systematic approach in the generation of the objects in the environment; namely, we generate their parameters through an equidistant sampling from the ranges given in Table \ref{Table:t1}. For example, in the angular domain, we assume a sector deployment with $120^{\circ}$ coverage from $30^{\circ}$ to $150^{\circ}$. Therefore, for three objects, the corresponding angles would be $[30,90,150]^{\circ}$, i.e., equidistantly spaced. For the Doppler domain, we do it slightly different: we also sample the relative velocities equidistantly from the range $[0, 25]\,$km/h, however, we set $50\%$ of them equal to zero. This is done to better reflect the fact that many of the paths seen in the environment correspond to static objects, e.g., buildings. 

Based on the objects, the channel is generated as a time-varying tapped delay line with multiple receive antennas. The \ac{OFDM} signal is then filtered with it, noise corrupted, and the received signal is demodulated and the channel tensor $\tensor{H}$ is estimated symbol-wise in the \ac{LS} sense. 

Since we evaluate the performance for deterministic scenarios, i.e., with fixed object parameters, we utilize here the average scenario-specific \ac{SNR} defined as
\begin{align}\label{eq:SNR}
	\textrm{SNR} = \frac{P_{\textsf{tx}}}{P_{\textsf{no}}}  \sum_{p=1}^{\Np} |\beta_p|^2\,,
\end{align}
where $P_{\textsf{tx}}$ and $P_{\textsf{no}}$ are the transmit and noise power, respectively. The noise power is calculated according to the given bandwidth and a temperature of $296\,$K, while the transmit power is adjusted based on the desired \ac{SNR} at the receiver.

Finally, in order to systematically compare the performance across different configurations and schemes, we use the \ac{RMSE} calculated as
\begin{align}
	\textrm{RMSE}_m = \sqrt{ \mathbb{E}\bigg\{\frac{1}{\Np} \sum_{p = 1}^{\Np} {|\phi_{m,p} - \hat{\phi}_{m,p}|}^2\bigg\}}\,,
\end{align}
where $\textrm{RMSE}_m$ is the \ac{RMSE} across the $m$-th estimation dimension. Here, how the error is calculated between the true and estimated parameters can be problematic. First, the detected parameters are not necessarily in the same order as the true parameters, e.g., the second object might be detected before the first object. To deal with this, we sort the true and estimated angles in an ascending order, and we calculate the \ac{RMSE} based on the sorted vectors. The exact sorting order is applied to the other dimensions (i.e., delay and Doppler). Using the same order is important to maintain correct pairing across the different dimensions. Second, if the true scenario has ten objects, while we detect only six, how to calculate the \ac{RMSE}? We handle this by selecting the true number of objects in the final selection step, when discussing \ac{RMSE} results.

\subsection{Role of Compression}
\begin{figure*}[!t]
    \centering
    \includegraphics[width=0.98\textwidth]{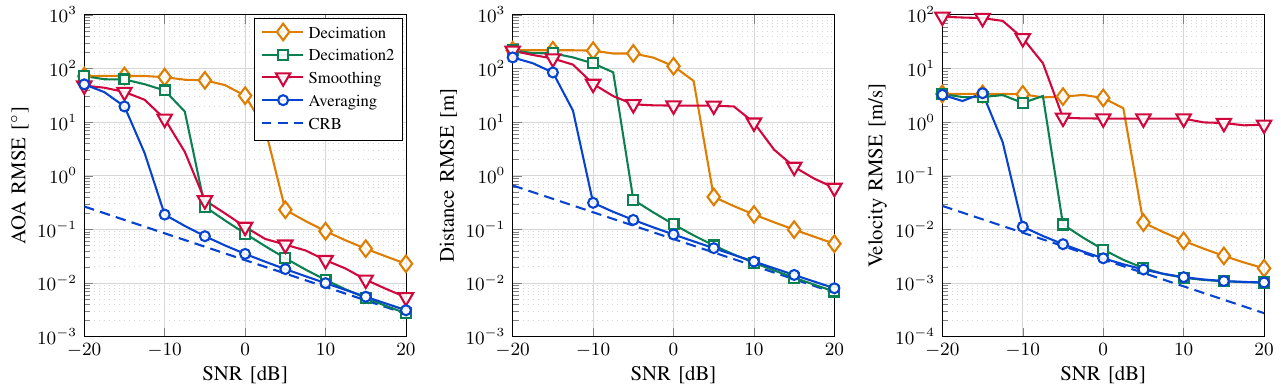}
    \captionsetup{justification=centering}
    \caption{\ac{RMSE} results for different compression methods for $\Np = 6$. Root-MUSIC is used with tensor-\ac{OMP} fusion. True $\Np$ is applied in the final selection.}
    \label{fig:res1}
\end{figure*}

The configuration of compression follows the same assumptions as those discussed in Section \ref{section:compression}, with a limit of $1$\,km on distance and $\pm 80\,$km/s on velocity. The resulting compression factors are shown in Table \ref{Table:t1}. We aim for maximum angular accuracy, and therefore we only perform compression over the frequency- and time-domains. Fig. \ref{fig:res1} shows the \ac{RMSE} results for the different compression schemes when $\Np = 6$ objects are present, with the corresponding \ac{CRB}.
During the estimation and fusion stages, the estimates $\hat{N}_{P,m}$ using \ac{AIC} over the decoupled modes are used. However, in the final selection step for $\mathcal{I}_{\textsf{Objects}}$, the true ${N}_{P}$ is used for correct \ac{RMSE} calculation.

For the decimation with virtual snapshots (``Decimation2'' in the figure), a total of $S = 56$ snapshots are generated. This is the maximum number of possible multidimensional shifts given the choice of the compression factors. As for smoothing, many more snapshots can be produced; in order to control the problem size, we limit the number of virtual snapshots to $S = 100$. These $100$ snapshots are sampled equidistantly over the possible multidimensional shifts of the smoothing operation.

As can be observed from the results, direct decimation causes a severe loss over \ac{SNR} as it throws away many samples of $\tensor{H}$. However, it maintains the high resolution capabilities, and it can be a useful low-complexity option, if high \ac{SNR} is guaranteed. The situation can be improved by considering other shifts of decimation as virtual snapshots, instead of throwing them away. These can be used to obtain a better estimate of the autocorrelation matrix used for root-MUSIC, and also later during fusion when estimating the core tensor. Smoothing, on the other hand, when applied in the context of decoupled estimation seems to severely affect the resolution of the smoothed dimensions. We observe an early saturation in the \ac{RMSE} results for both the distance and velocity estimation, due to decrease in aperture. Finally, averaging seems to be the perfect choice in our considered scenario. On the one hand, it produces an \ac{SNR} gain due to averaging of neighboring samples. On the other hand, it does not produce additional snapshots, and therefore the estimation and fusion stages will operate with a single snapshot only, thereby decreasing complexity. We therefore adopt averaging in the following.

\subsection{Influence of Fusion Method}
\begin{figure*}[!t]
    \centering
    \includegraphics[width=0.98\textwidth]{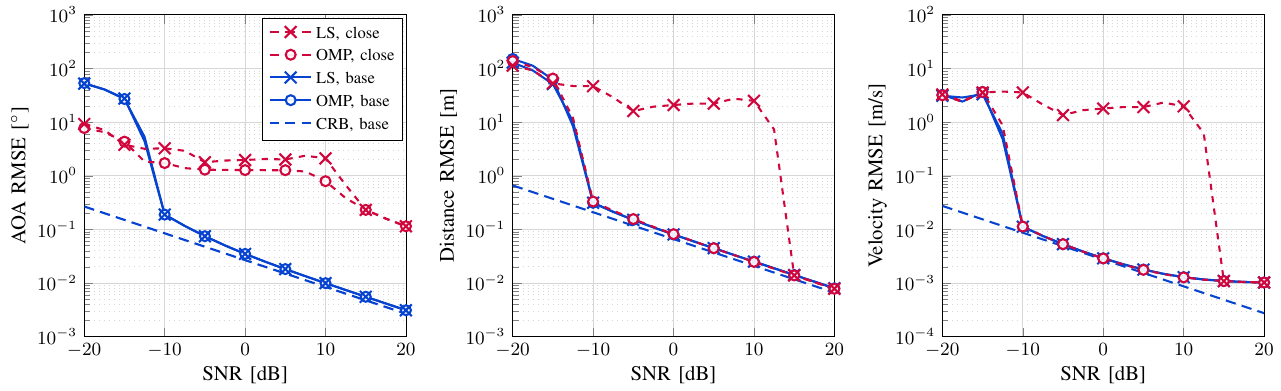}
    \captionsetup{justification=centering}
    \caption{\ac{RMSE} results for \ac{LS}- and \ac{OMP}-based fusion for $\Np = 6$. Averaging with root-MUSIC is used. True $\Np$ is applied in the final selection.}
    \label{fig:res2}
\end{figure*}

We also investigate if the choice of the fusion method plays a major role in how the parameters are paired together. We investigate this for two setups using $\Np = 6$ objects in Fig. \ref{fig:res2}. The first setup follows the same procedure as before with the ``base'' scenario. The second setup, which we refer to as ``close'', places the objects closely-spaced in the angular domain. More specially, the angles range is restricted to $[80, 100]^{\circ}$, instead of the original $120^{\circ}$ spread. The ``close'' case then provides a scenario where the parameters are more difficult to resolve. In order to isolate the possible effect of high angular errors on sorting for the \ac{RMSE} calculation, we do the \ac{RMSE} sorting in this example based on distance.

For the ``base'' case, we observe full overlap between the \ac{LS}- and \ac{OMP}-based fusion. This is actually very promising, since it allows us to perform the fusion as one-shot per-dimension \ac{LS}-based operations. As for the ``close'' scenario, we see a different picture. In the angular \ac{RMSE}, we see a worsened performance compared to the base case; this is expected since the objects now share very similar angles, and therefore fully resolving them becomes an issue. In the distance and velocity results, we see a huge gap between \ac{LS}- and \ac{OMP}-based fusion. We see that tensor-\ac{OMP} provides almost the same performance as that of the base setup. However, tensor-\ac{LS} falls short in that regard and is only able to correctly fuse the parameters under high \ac{SNR}. This is an indicator that for scenarios with closely-spaced objects, whether in angles, distances, and/or velocities, utilizing the sparse structure of the core tensor using an \ac{OMP}-based scheme, can provide a more robust implementation of the fusion/pairing stage.

\subsection{Comparison to Other Schemes}
For comparing the performance against other schemes in the literature, we opted here for a generic comparison focusing on the main structure of two approaches: sequential and joint. Due to the large number of variants available, we focus here on discussing the general advantages and disadvantages of our scheme compared to these categories, rather than focusing on a particular scheme with a certain configuration or approach.

\subsubsection{Sequential Baseline}
No form of compression is performed here. At the start, the angles are estimated using root-MUSIC, by treating other dimensions as the snapshots dimension. For each detected angle, 2D-\ac{DFT} processing is performed, transforming the time-frequency plane to the delay-Doppler plane. The largest delay-Doppler peak with the associated angle is then declared to be an active object. To clarify further, assuming the estimated angles response matrix to be $\hat{\mathbf{A}} \in \mathbb{C}^{\Na \times \hat{N}_{P}}$, we construct the time-frequency view
\begin{align}
	\tensor{H}_{\textsf{TF}} = \tensor{H} \times_1 \hat{\mathbf{A}}^{\dag}\,,
\end{align}
where each slice of $\tensor{H}_{\textsf{TF}} \in \mathbb{C}^{\hat{N}_{P} \times \Ns \times \Nt}$ across its first dimension is a time-frequency view along one of the detected angles. By performing 2D-\ac{DFT} processing over each of these slices, we get to the final results. 

\subsubsection{Joint Baseline}
For the joint baseline, we implement here the Tensor-ESPRIT algorithm \cite{Haardt08}.  
Since Tensor-ESPRIT requires the construction of a global signal subspace through a \ac{HOSVD} \cite{Cichocki15}, we use smoothing, as done in the literature, to construct virtual snapshots for that purpose. In order to limit the influence of smoothing on resolution, we limit smoothing to a factor of two over frequency and time, i.e., $\Delta_2 = \Delta_3 = 2$. Otherwise, it can severely affect the estimation performance. This of course results in higher estimation complexity compared to our architecture, which uses the factors listed in Table \ref{Table:t1}. 

For the construction of signal subspaces across the different modes, we rely on a parallel \ac{HOSVD} construction, instead of a serial one, in order to limit the error propagation from one subspace to another, if the estimated number of components is incorrect. 
For each mode, the number of components are determined separately using its corresponding \ac{BIC}. We use \ac{BIC} here instead of \ac{AIC} due to the large sample size resulting from smoothing, allowing for better performance. In the final selection stage, where the global signal subspace is constructed, we use the true $\Np$, following the same reasoning as before with respect to the calculation of the \ac{RMSE}. For the joint diagonalization step in Tensor-ESPRIT, we apply a variant of the \ac{RJD} method proposed in \cite{He24}. In our tests, it could stably provide joint basis in less than five iterations.

\begin{figure*}[!t]
    \centering
    \includegraphics[width=0.98\textwidth]{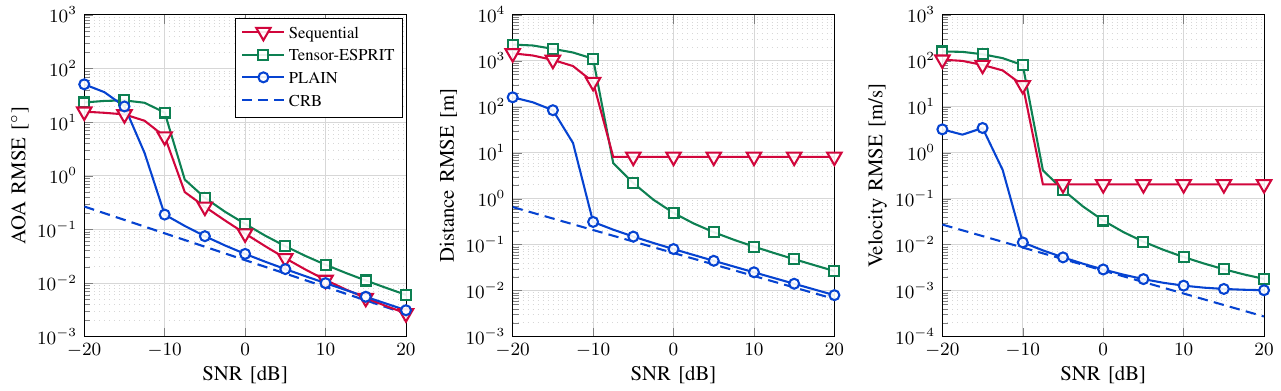}
    \captionsetup{justification=centering}
    \vspace{-0.5em}
    \caption{\ac{RMSE} results for different estimation schemes for $\Np = 6$. Averaging with root-MUSIC is used for \ac{PLAIN}. True $\Np$ is applied in the final selection.}
    \label{fig:res3}
\end{figure*}

Fig. \ref{fig:res3} shows the \ac{RMSE} results of a scenario with six objects using our proposed approach with averaging and tensor-\ac{OMP} compared to the generic sequential scheme and Tensor-ESPRIT.
In the angular domain, we see that all the schemes are capable of providing high resolution approaching what is theoretically predicted by the \ac{CRB}. The sequential scheme with no compression seems to even outperform Tensor-ESPRIT at high \ac{SNR}. This is due to the usage of root-MUSIC algorithm, which has a slight upper-hand compared to ESPRIT. For the distance and velocity results, we see a saturation of the sequential scheme. This is due to the usage of the 2D-\ac{DFT} for delay-Doppler estimation, which depends on the resolution of the \ac{DFT} bins. Tensor-ESPRIT does not suffer from the resolution limitation, and is able to provide high resolution scalability at high \ac{SNR}. Our proposed architecture provides super resolution capabilities already in the low \ac{SNR} range, while maintaining low processing complexity, since it only deals with a single snapshot generated by averaging.

\begin{figure*}[!t]
    \centering
    \includegraphics[width=0.98\textwidth]{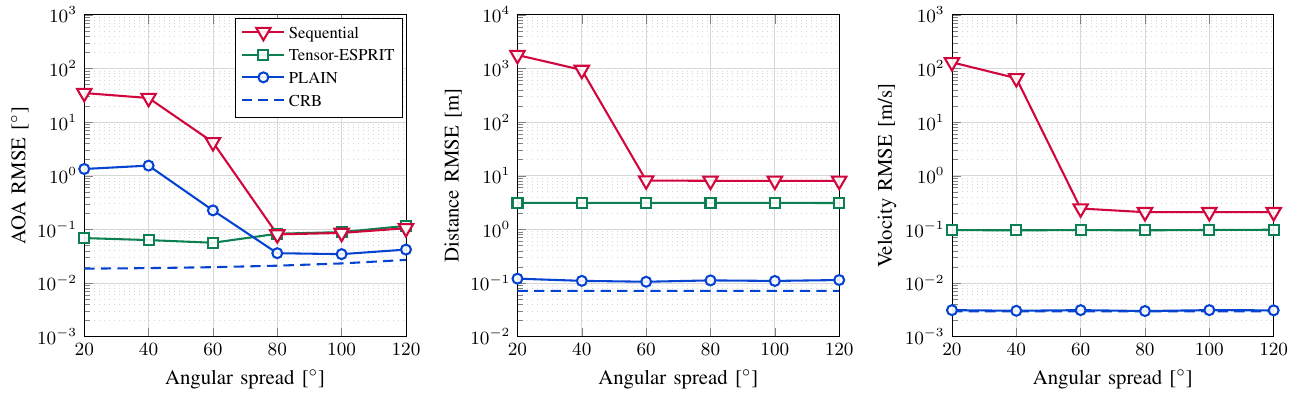}
    \captionsetup{justification=centering}
    \vspace{-0.8em}
    \caption{\ac{RMSE} results for $\Np = 10$ objects for various angular spacings. Scenario with $\textrm{SNR} = 0\,$dB. True $\Np$ is applied in the final selection.}
    \label{fig:res4}
\end{figure*}

Next, we investigate the influence of object spacing on the performance of these approaches. We consider now a scenario with $\Np = 10$ objects and evaluate the performance over different angular ranges around $90^{\circ}$, starting from the range of $[80,100]^{\circ}$ ($20^{\circ}$ spread) to the full $[30,150]^{\circ}$ range ($120^{\circ}$ spread). This is shown in Fig. \ref{fig:res4}, where the scenario \ac{SNR} is fixed to $0\,$dB.

When the objects are sufficiently spaced far from each other, we observe that all schemes are capable of resolving the objects and providing good accuracy. When the objects are close, this starts to have an impact on the estimation performance. This is the worst for the sequential scheme, since once an object is missed in the angular search, it becomes unrecoverable, which is propagated to the following distance and velocity estimation. For the other schemes, due to the multidimensional combining, information from multiple dimensions are used to infer the activity of objects. For our proposed scheme, we see once the angular spread becomes narrow, then worse performance is obtained. However, this does not impact the estimation accuracy across the other dimensions, as in the sequential scheme. Moreover, due to the fusion with the other dimensions, the angular performance does not deteriorate substantially, but rather saturates around the limited resolution of \ac{1D} estimation. Specifically, single-digit degree accuracy can still be achieved. For the Tensor-ESPRIT algorithm, it maintains the high accuracy even for tight spacing. This is due to the joint estimation of the signal space across the different tensor modes, which gives it an upper-hand compared to our approach, however, at a higher processing complexity.

\begin{figure}[!t]
    \centering
    \includegraphics[width=0.98\linewidth]{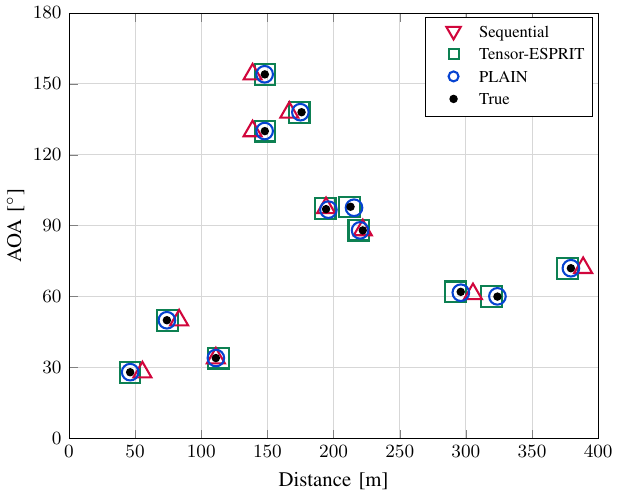}
    \captionsetup{justification=centering}
    \vspace{-0.8em}
    \caption{Angle-distance view for an arbitrary deployment with $\textrm{SNR} = 0\,$dB. Estimated number of objects $\hat{N}_P$ is applied the final selection stage as well.}
    \label{fig:res5}
\end{figure}

Furthermore, we investigate an arbitrarily scenario with $\Np = 12$ objects, as shown in Fig. \ref{fig:res5}. 
Here, the estimated $\hat{N}_P$ is also used in the final selection process, since we do not need to calculate an \ac{RMSE}. We make similar observations as before; for closely-spaced objects, the sequential scheme fails to resolve them properly. This can be clearly seen for the objects around the distances of $200\,$m and $300\,$m. The tensor-based schemes, on the other hand, show robustness to the tight spacing and are able to better identify the objects. 

Finally, we measure the runtime of the three algorithms on the scenario in Fig. \ref{fig:res5}. This is listed in Table \ref{Table:t2} in seconds for an average of 100 runs. To get a sense of scalability, we also list a ``doubled setup'', where the number of antennas and bandwidth are doubled to $\Na = 32$ antennas and $21.6\,$MHz, respectively. As can be observed, the proposed \ac{PLAIN} architecture can finish at a time order similar to that of the sequential scheme, while running super-resolution algorithms across all the dimensions. 
For Tensor-ESPRIT, the calculations required for the \ac{HOSVD}, in addition to the \ac{LS} problems to be solved and also smoothing, seem to cause long processing time. This is especially pronounced in the doubled setup, taking around $14\,$s to finish. One way to improve the situation with Tensor-ESPRIT is to increase the smoothing factors; however, as discussed before, this can severely affect the resolution. Another way is to take less virtual snapshots from smoothing; this can help with reducing the time spent on the \ac{HOSVD}, at the cost of a less accurate estimation of the global signal space. Overall, we observe that the \ac{PLAIN} architecture, with its single-snapshot high-resolution support and accommodation for high dimensionality, provides an attractive choice for addressing complexity, scalability, and super-resolution.

\begin{table}[t]
    \begin{center}
        \small
        \def\arraystretch{1.1}
        \caption{Average runtime out of 100 runs for the scenario in Fig. \ref{fig:res5}.}
        \label{Table:t2} 
        \begin{tabularx}{0.9\linewidth}{l | c |c |c}
            & Sequential  & Tensor-ESPRIT & {\ac{PLAIN}} \\ \hline
            Base setup & $0.04\,$s & $3.53\,$s & $0.05\,$s \\
            Doubled setup & $0.10\,$s & $14.05\,$s & $0.22\,$s \\
            \hline
        \end{tabularx}
    \end{center}
\end{table}

\section{Conclusion}
In this work, we proposed \ac{PLAIN}, a tensor-based estimation architecture that can handle a high number of sensing dimensions, high dimensionality, limited sensing time, while operating with low-complexity and providing super-resolution capabilities. The architecture addresses dimensionality through a compression stage that reduces the problem size, while persevering information required to achieve high resolution. The existence of many sensing dimensions is handled through a parallel decoupled estimation stage, where each decoupled branch only deals with estimating the parameters related to the corresponding estimation dimension. Finally, the parameters are fused together to form a multidimensional estimate, by exploiting the tensor and sparse structure of the problem. We investigated different compression and fusion methods, and compared the performance against a generic sequential baseline, as well as the multidimensional Tensor-ESPRIT algorithm. The results show that PLAIN can strike a good balance between complexity and performance, and allows for a highly flexible estimation structure that can be tailored to various deployments and sensing formats.

\input{./config/acronyms.tex}
\bibliographystyle{IEEEtran}
\bibliography{IEEEabrv,./config/references}

\end{document}

%% file: config/acronyms.tex
\begin{acronym}[DSTTDSGRC]
\setlength{\itemsep}{-3pt}
\acro{CS}{compressed sensing}
\acro{DFT}{discrete Fourier transform}
\acro{MMSE}{minimum mean squared error}
\acro{SNR}{signal-to-noise ratio}
\acro{LTE}{long-term evolution}
\acro{SINR}{signal-to-interference-plus-noise ratio}
\acro{BLER}{block error ratio}
\acro{4G}{fourth-generation}
\acro{5G}{fifth-generation}
\acro{6G}{sixth-generation}
\acro{B5G}{beyond fifth-generation}
\acro{IoT}{internet-of-things}
\acro{FFT}{fast-Fourier-transform}
\acro{IFFT}{inverse fast-Fourier-transform}
\acro{OFDM}{orthogonal frequency-division multiplexing}
\acro{MMSE}{minimum mean square error}
\acro{MF}{matched filter}
\acro{RB}{resource-block}
\acrodefplural{RB}{resource-blocks}
\acro{RMS}{root-mean-square}
\acro{MIMO}{multiple-input multiple-output}
\acro{V2X}{vehicle-to-everything}
\acro{IEEE}{institute of electrical and electronics engineers}
\acro{MAC}{medium access control}
\acro{PHY}{physical layer}
\acro{5G-NR}{5G new-radio}
\acro{IRS}{intelligent reflecting surface}
\acro{RIS}{reconfigurable intelligent surface}
\acro{LOS}{line-of-sight}
\acro{NLOS}{non-line-of-sight}
\acro{ISAC}{integrated sensing and communication}
\acro{MUSIC}{MUltiple SIgnal Classification}
\acro{RF}{radio frequency}
\acro{KPI}{key performance indicator}
\acro{AOA}{angle of arrival}
\acro{BS}{base station}
\acro{UE}{user equipment}
\acro{CSI}{channel state information}
\acro{ULA}{uniform linear array}
\acro{RCS}{radar cross-section}
\acro{PL}{path loss}
\acro{PLAIN}{com\dunderline{p}ressed decoup\dunderline{l}ed estim\dunderline{a}tion and \dunderline{i}nput-based fusio\dunderline{n}}
\acro{1D}{one-dimensional}
\acro{mmWave}{millimeter wave}
\acro{FBA}{forward-backward averaging}
\acro{AIC}{Akaike information criterion}
\acro{BIC}{Bayesian information criterion}
\acro{LS}{least-squares}
\acro{OMP}{orthogonal matching pursuit}
\acro{CRB}{Cram\'er-Rao bound}
\acro{CP}{cyclic-prefix}
\acro{RMSE}{root mean squared error}
\acro{HOSVD}{higher-order singular value decomposition}
\acro{RJD}{randomized joint diagonalization}
\acro{CPD}{canonical polyadic decomposition}

\end{acronym}